\documentclass[]{emulateapj}    
\usepackage{graphicx} 
\usepackage{epstopdf} 
\usepackage{longtable} 
\usepackage{natbib}
\usepackage{chngpage}

\begin{document}

\newcommand{\met}{[Z]}
\newcommand{\jband}{J\textendash{}band}
\newcommand{\logg}{log$g$~}
\newcommand{\csq}{$\chi^2$~}
\newcommand{\teff}{ T$_{\rm{eff}}$~}
\newcommand{\lte}{LTE}
\newcommand{\nlte}{NLTE}
\newcommand{\hii}{H\,{\sc ii}\rm}

\def\Msun{\hbox{M$_\odot$}}
\newcommand{\degree}{\hbox{$^\circ$$$}}
\newcommand{\decdeg}{\hbox{$.\!\!^\circ$}}
\newcommand{\darcmin}{\hbox{$.\mkern-4mu^\prime$}}
\newcommand{\darcsec}{\hbox{$.\!\!^{\prime\prime}$}}
\newcommand{\hour}{\hbox{$^{\rm h}$}}
\newcommand{\dech}{\hbox{$.\!\!^{\rm h}$}}
\newcommand{\minute}{\hbox{$^{\rm m}$}}
\newcommand{\decmin}{\hbox{$.\!\!^{\rm m}$}}
\newcommand{\secnd}{\hbox{$^{\rm s}$}}
\newcommand{\decsec}{\hbox{$~\!\!^{\rm s}$}}
\newcommand{\RA}[4]{#1\hour#2\minute#3\decsec}            
\newcommand{\DEC}[3]{$#1$\degree$#2$\arcmin$#3$\arcsec}      

\title{A New Method for Measuring Metallicities of Young Super Star Clusters}
\author{
J. Zachary Gazak\altaffilmark{1,2},
Ben Davies\altaffilmark{3}, 
Nate Bastian\altaffilmark{3},
Rolf Kudritzki\altaffilmark{1,4},
Maria Bergemann\altaffilmark{5},
Bertrand Plez\altaffilmark{6},
Chris Evans\altaffilmark{7,8},
Lee Patrick\altaffilmark{8},
Fabio Bresolin\altaffilmark{1},
Eva Schinnerer\altaffilmark{9}
}

\bibliographystyle{apj}  

\begin{abstract}
We demonstrate how the metallicities of young super star clusters can be measured using novel spectroscopic 
techniques in the \jband.  The near-infrared flux of super star clusters older than $\sim$6 Myr is dominated by 
tens to hundreds of red supergiant stars.  Our technique is designed to harness the integrated light of that 
population and produces accurate metallicities for new observations in galaxies above (M83) and below 
(NGC 6946) solar metallicity.  In M83 we find \met\ = $+$0.28 $\pm$ 0.14 dex using a moderate resolution 
 (R$\sim$3500) \jband\ spectrum and in NGC 6496 we report \met\ = \textendash0.32 $\pm$ 0.20 dex from 
a low resolution spectrum of R$\sim$1800.  
Recently commissioned low resolution multiplexed spectrographs on
the VLT (KMOS) and Keck (MOSFIRE) will allow accurate measurements of super star cluster metallicities 
across the disks of star-forming galaxies up to distances of 70 Mpc with single night observation campaigns 
using the method presented in this letter.
\end{abstract}


\altaffiltext{1}{Institute for Astronomy, University of
Hawai'i, 2680 Woodlawn Dr, Honolulu, HI 96822, USA}
\altaffiltext{2}{Visiting Astronomer at the Infrared Telescope Facility, which is operated by the University of 
Hawaii under Cooperative Agreement no. NNX-08AE38A with the National Aeronautics and Space 
Administration, Science Mission Directorate, Planetary Astronomy Program.}
\altaffiltext{3}{Astrophysics Research Institute, Liverpool John Moores University, 146 Brownlow Hill, Liverpool L3 5RF, UK}
\altaffiltext{4}{University Observatory Munich, Scheinerstr. 1, D-81679 Munich, Germany}
\altaffiltext{5}{Institute of Astronomy, University of Cambridge, Madingley Road, Cambridge CB3 0HA, UK}
\altaffiltext{6}{Laboratoire Univers et Particules de Montpellier, Universit\'e Montpellier 2, CNRS, F-34095 Montpellier, France}
\altaffiltext{7}{UK Astronomy Technology Centre, Royal Observatory Edinburgh, Blackford Hill, Edinburgh., EH9 3HJ, UK}
\altaffiltext{8}{Institute for Astronomy, Royal Observatory Edinburgh, Blackford Hill, Edinburgh., EH9 3HJ, UK}
\altaffiltext{9}{MPI for Astronomy, Konigstuhl 17, 69117 Heidelberg, Germany}

\maketitle

\section{Introduction}
\label{sec:intro}


The effects of star formation$-$notably the chemical enrichment of a galaxy's 
young stellar population and interstellar medium$-$imprint a signature of the initial properties and evolution 
of that galaxy onto its current generation of stars.  Two critical observables are the 
central metallicity and radial abundance gradient of iron- and $\alpha$-group 
elements.  Trends in such measurements over ranges of galactic mass, redshift,
and environment constrain the theory of galaxy formation and chemical evolution.

The central metallicity of a galaxy is dictated by mass and traces formation properties
and evolution \citep{1979AA....80..155L,2004ApJ...613..898T,2008AA...488..463M}.
Radial abundance gradients provide a signature of the complex dynamics of galaxy
evolution and the growth of galactic disks.  The processes affecting abundance gradients include clustering, merging,
infall, galactic winds, star formation history, and initial mass function \citep{2000MNRAS.313..338P,2004cmpe.conf..171G,2008AA...483..401C,2009AA...505..497Y,2009MNRAS.398..591S,2004MNRAS.349.1101D,2007MNRAS.374..323D,2008MNRAS.385.2181F,2007ApJ...655L..17B,2007MNRAS.375..673K,2009MNRAS.399..574W}.  

Pursuit of these rich areas of research has been undermined by the difficulty of obtaining 
reliable metallicities of galaxies.  Techniques providing such measurements must be observationally 
efficient as well as accurate and precise.  These requirements offer a formidable challenge 
across extragalactic distances.  In the ideal situation, conclusions are drawn from careful 
studies using multiple techniques such as described below.  Constraints on access to telescopes and the lack of 
targets which can be observed over such distances necessitates compromise.

The bulk of investigations rely on spectroscopy of the emission lines of \hii\ regions.  
The ``strong line'' analysis methods use the fluxes of the strongest forbidden lines 
relative to H$\beta$.  The requirement for empirical calibration has created a situation 
in which different commonly used calibrations yield varying and sometimes conflicting 
results from the same set of observations.  Both the slope and absolute scaling of 
metallicity are susceptible to choice of calibration: the mass-metallicity gradient across 
all galaxies and the radial gradients within individual galaxies can change from steep 
to flat while the overall metallicity can shift by a factor of up to four 
\citep{2008ApJ...681.1183K,2008ApJ...681..269K,2009ApJ...700..309B}.  Even the more 
physical ``T$_e-$based method'' (which utilizes auroral lines to remove the need for 
``strong line'' calibrations) is potentially subject to biases$-$especially in the metal rich 
regime characteristic of the disks of all massive spiral galaxies 
\citep{2014arXiv1401.4437B,2005AA...434..507S,2005AA...441..981B,2010MNRAS.401.1375E,2012MNRAS.427.1463Z}.

\begin{figure*}
\begin{centering}
\includegraphics[width=7in]{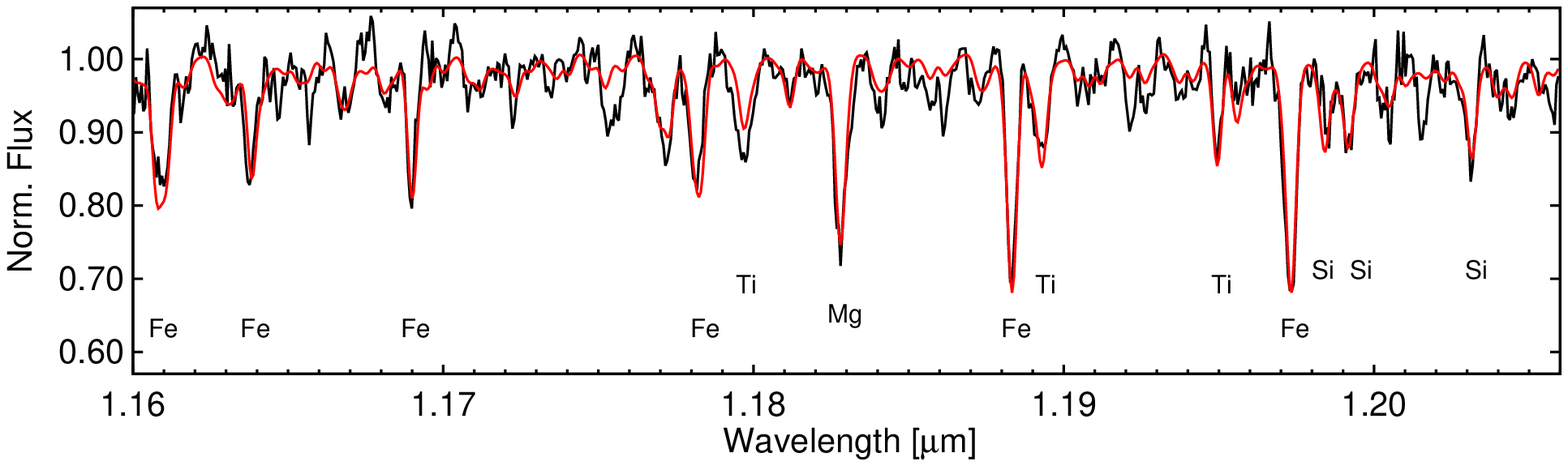}
\includegraphics[width=7in]{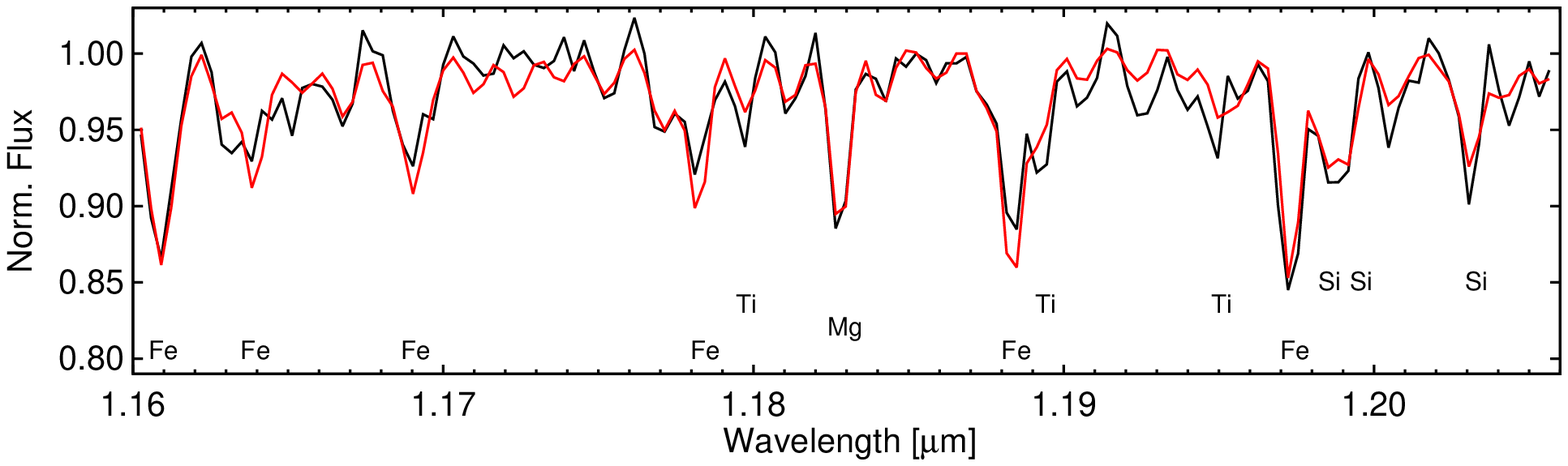}
\caption{Observed spectra of two young super star clusters (black): M83-1f-117 (upper panel), and NGC6946-1447 (lower panel).  Each spectrum is overplotted with a best fitting red supergiant synthetic spectrum (red dashed with gray) in the spectral window analyzed.  The critical diagnostic lines of Fe\,{\sc i}, Ti\,{\sc i}, Si\,{\sc i}, and Mg\,{\sc i} are marked.  Spectral fitting is carried out over these diagnostic features.} 
\label{fig:specfit}
\end{centering}
\end{figure*}

Quantitative spectroscopy of supergiant stars is one alternative technique which
avoids the uncertain calibrations of the ``strong line'' \hii\ region method.  Blue 
supergiants in particular have become a powerful tool for measuring metallicities, abundance gradients, 
and distances to galaxies in and beyond the Local Group (WLM \textendash{} 
\citealt{2006ApJ...648.1007B,2008ApJ...684..118U}; NGC 3109 \textendash{} 
\citealt{2007ApJ...659.1198E}; IC1613 \textendash{} \citealt{2007ApJ...671.2028B}; 
M33 \textendash{} \citealt{2009ApJ...704.1120U}; 
M81 \textendash{} \citealt{2012ApJ...747...15K}; NGC 4258 \textendash{} \citealt{2013ApJ...779L..20K}).  
This technique, while extremely promising, may also be subject to systematic 
uncertainties and needs to be checked by independent methods. Moreover, it utilizes 
optical spectroscopy, while the next generation of telescopes such as the 
TMT and  E\textendash{}ELT will be optimized for observations using adaptive
optics at infrared 
wavelengths (IR).  Bright abundance tracers with their flux-maximum in the
IR will have a clear advantage as these facilities come online.

Red supergiants (RSGs) are extremely luminous stars which emit 
10$^5$ to $\sim$10$^6$ L/L$_\odot$ largely in the infrared 
\citep{Humphreys:1979p3252}.  With a method to extract
metallicities from these stars at modest resolutions the RSGs become 
ideal targets for measuring extragalactic cosmic abundances.  Studies 
of RSGs tend to demand high spectral resolutions (R$\approx$20,000) 
in order to disentangle the densely packed atomic and 
molecular features iconic to such cool, inflated stars.  By searching for
a spectral window with minimal contamination by the strongest molecular
lines of OH, H$_2$O, CN, and CO,  \cite{2010MNRAS.407.1203D} found 
the \jband\ (1.15-1.23 ${\mu}$m) a suitable bandpass.  Furthermore, the 
dominant spectral features are isolated atomic lines of iron, titanium, 
silicon, and magnesium.  \cite{2010MNRAS.407.1203D} and, most recently \cite{gazak2014} have 
demonstrated the extraction of metallicities accurate to $\sim$0.10 dex for a single RSG at 
resolutions down to R$\approx$2,000 in the \jband.

With this \jband\ method in hand, current instrumentation on 8-meter class 
telescopes can extract accurate and precise metallicities using single
RSGs out to a limiting distance of $\sim$10 Mpc \citep{2011AA...527A..50E}.  
Still, to reach groups and clusters of galaxies significantly beyond the 
10 Mpc, such techniques must await the next generation of 30-meter
class telescopes.  

In this work we demonstrate a new method for extending the observational
baselines of stellar techniques  by exploiting the integrated light of coeval 
ensembles of stars.  In star forming galaxies, such populations exist as super 
star clusters (SSCs), the result of single bursts of star formation creating a 
population with a stellar mass of 10$^4$-10$^6$ M$_{\odot}$ in a tight 
association \citep{2010ARAA..48..431P}.  In \cite{2013MNRAS.430L..35G} 
we hypothesized that a discrete jump in the IR colors of SSCs at ages beyond
7 Myrs was caused by the appearance and flux dominance of the RSG 
members of these clusters.  We demonstrated this to be the case by performing
population synthesis experiments with synthetic photometry.  The simulations
agreed well with observed near IR colors of SSCs across a range of ages in
M83 measured by \cite{2011MNRAS.417L...6B,2012MNRAS.419.2606B}.  Indeed, by 7 Myr 
the population of tens to hundreds of RSGs dominates the near-IR light, commanding
 $\ge$ 90-95\% of the \jband\ flux \citep{2013MNRAS.430L..35G}.  
As a natural extension of that work we suggested that SSCs older than 7 Myr
could be used for quantitative spectroscopy and the measurement of \met\ at 
far greater distances than is possible for single supergiants.  It is the purpose
of this paper to demonstrate the practicality of this spectroscopic technique and
present applications significantly above and below solar metallicity.

The analysis method applied to the \jband\ spectra is the same as used very recently
in the quantitative spectroscopic \jband\ study of individual RSGs in the Milky Way
cluster Perseus OB-1 by \cite{gazak2014}.

This novel method allows for the measurement of metallicities of young super star 
clusters$-$and thus the disks of star-forming galaxies$-$within $\sim$35 Mpc 
and across a wide range of galactic metallicity.  To this end we have collected 
\jband\ spectra of two SSCs, one in the disk of the super-solar metallicity galaxy 
NGC5236 (M83) at 4.5 Mpc \citep{2003ApJ...590..256T} and one in the sub-solar metallicity galaxy NGC 
6946 at 5.9 Mpc \citep{2000AA...362..544K}.  This represents the pioneering first step towards studying 
the disks of star-forming galaxies with stellar spectroscopy over distances extending 
ten times that of single-supergiant techniques.  To accomplish this 
we observed M83$-$1f$-$117 (referred to as NGC5236-805 in 
\citealt{2004AA...427..495L}), a m$_{\rm{J}}$=16.1 SSC at an age of $\sim$20 Myr 
and mass of 2$\times$10$^5$ M$_\odot$ in the nearby spiral galaxy M83.  For 
the sub-solar case we targeted NGC 6946-1447, a m$_{\rm{J}} \sim$13 SSC at 
an age of $\sim$10-15 Myr with a mass of $\sim$10$^{6}$ M$_\odot$.

\begin{figure*}
\begin{centering}
\includegraphics[width=8.8cm]{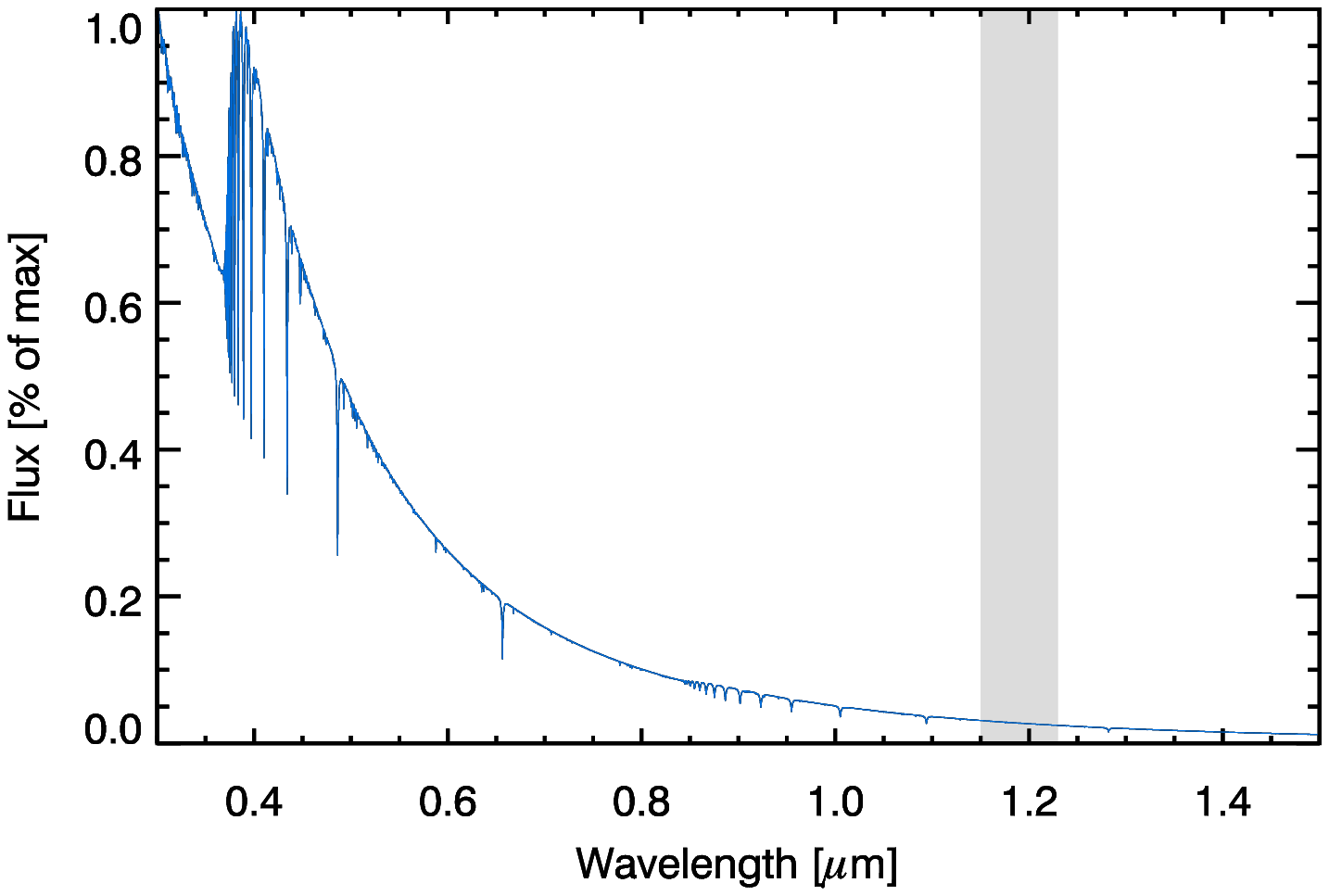}
\includegraphics[width=8.8cm]{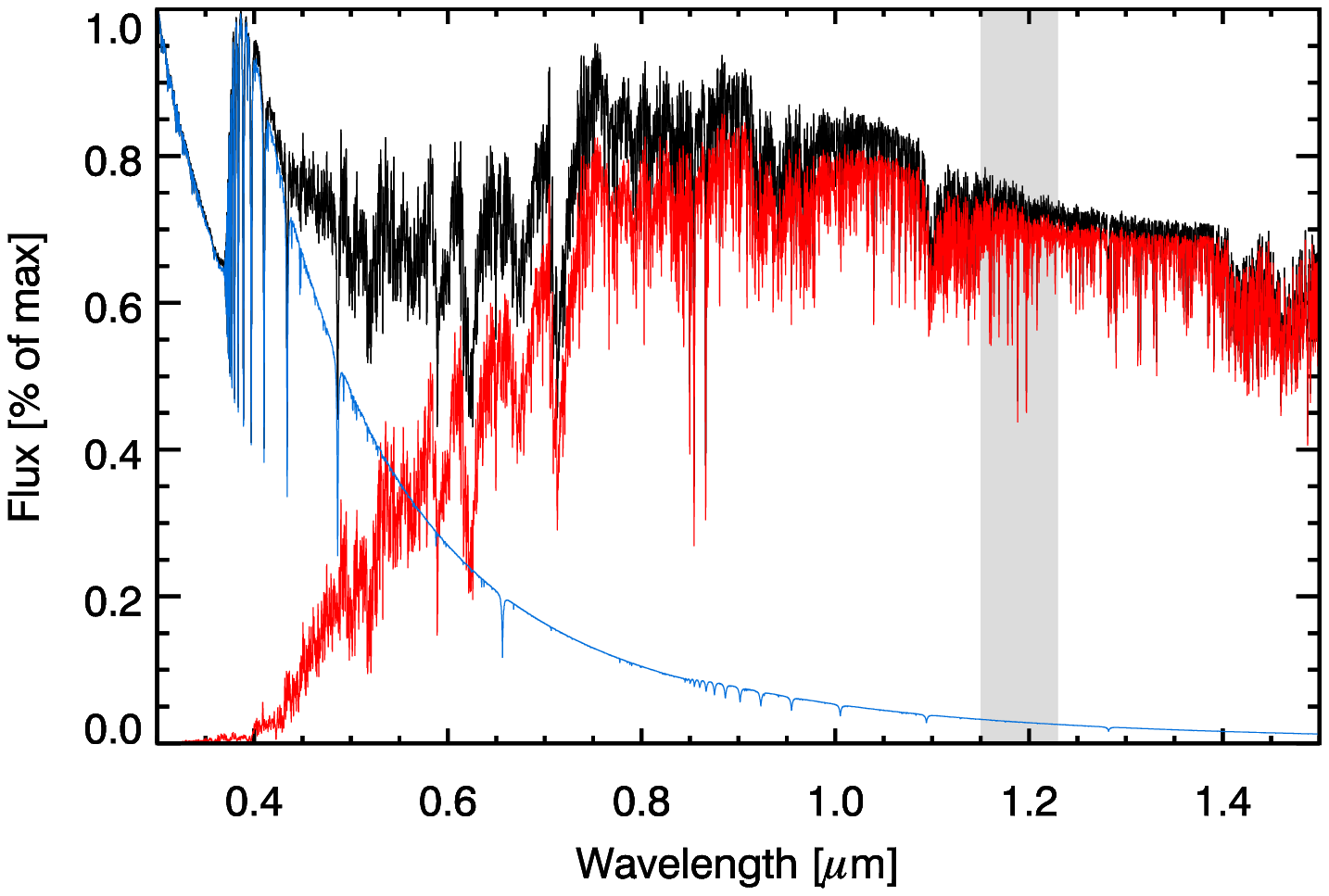}
\caption{Theoretical spectra for a 10$^5$ M$_\odot$ super star cluster after 5 Myr (left panel) and 15 Myr (right panel).  Black spectrum represents the full SSC SED, blue represents the main sequence and blue supergiant stars, and red plots flux due to red supergiant members.  The \jband\ is highlighted in gray.  } 
\label{fig:syn1}
\end{centering}
\end{figure*} 

\section{Observations}
\label{sec:obs}

Observations of M83-1f-117 ($\alpha$=\RA{13}{37}{02}{0}, $\delta$=\DEC{-29}{52}{13}) 
were obtained using ISAAC/VLT  \citep{1998Msngr..94....7M} on the night of 2012 March 13 
under the ESO programme 089.D-0750(A) (P.I. Bastian, N).  We employed the 
1$\arcsec$.0 slit width with a central wavelength of 1.17 $\mu$m and integrated 
on source for two hours using an ABA nod pattern.  We observed a B-type star with 
a similar airmass as a telluric standard. 

The spectra were reduced following the methodology outlined in \cite{2012MNRAS.419.1871D}.  
Briefly, this reduction consists of the subtraction of nod pairs, flat-fielding, rectification to 
correct for distortion in the spatial and dispersion directions, sky subtraction, and cosmic-ray removal.

NGC6946-1447 ($\alpha$=\RA{20}{34}{52}{0}, $\delta$=\DEC{60}{08}{14}) was observed 
on 2011 August 3 and 2011 October 12 with the near-IR medium resolution SpeX spectrograph 
mounted on the 3$-$meter NASA InfraRed Telescope Facility (IRTF) on the summit of Mauna 
Kea \citep{2003PASP..115..362R}.  SpeX was set up in short wavelength cross-dispersed 
mode with a 0$\arcsec$.3 slit.  The data were reduced and telluric-corrected using the IDL 
spectral extraction package Spextool \citep{2003PASP..115..389V,2004PASP..116..362C}.

The observed spectra are plotted in Figure \ref{fig:specfit}.

\begin{deluxetable}{llcccc}
\centering
\tabletypesize{\small}
\tablewidth{0pt}
\tablecaption{Model Grid}
\tablehead{
\colhead{Param} &\colhead{} &  \colhead{Grid Min}   & \colhead{Grid Max}  & \colhead{Spacing}}
\startdata 
 \teff    & [K]              & 3400  & 4000  & 100 \\
              &             &  4000  & 4400 & 200 \\
 log$g$ &  [cm s$^{-2}$]    & $-$1.0  & +1.0  &  0.5 \\
 $\rm{\met}$\tablenotemark{a}  &    & $-$1.00 & +1.00  & 0.25 \\
 $\xi$&  [km s$^{-1}$]             & 1.0  & 6.0   &  1.0 
\enddata
\tablecomments{The parameter space of the {\sc marcs} grid of stellar atmospheres used in this work.}
\label{tbl:par_grid}
\end{deluxetable}


\begin{deluxetable}{lccc}
\centering
\tabletypesize{\small}
\tablewidth{0pt}
\tablecaption{Spectral fits}
\tablehead{
\colhead{Param}  & \colhead{M81-1f-117} &  \colhead{NGC6946-1447} }
\startdata 
 \teff            &      3540 $\pm$ 80         & 3940 $\pm$ 100   \\
 log$g$      & $+$0.48 $\pm$ 0.18    & $+$0.10 $\pm$ 0.15  \\
 $\rm{\met}$\tablenotemark{a}    &  $+$0.28 $\pm$ 0.14   & $-$0.32 $\pm$ 0.20  \\
 $\xi$         &      3.1 $\pm$ 0.25        & 3.0 $\pm$ 0.25   \\
R$_{\rm{eff}}$  [$\lambda/\delta\lambda$]  & 3500 $\pm$ 50 & 1800 $\pm$ 50
\enddata
\tablecomments{Parameter fits to the observed spectra.}
\label{tbl:par_grid2}
\end{deluxetable}

\section{Synthetic Super Star Clusters}

\subsection{Stellar Population Synthesis}

\cite{gazak2014}, in their analysis of \jband\ spectra of individual RSGs in the Milky Way cluster Perseus OB-1, synthesized an integrated \jband\ cluster spectrum by combining the individual spectra of all RSGs studied.  They demonstrate that the quantitative analysis of the integrated spectrum yields a metallicity consistent to the mean metallicity of the individual RSGs.  This is the starting point for the population synthesis experiment described in this section.  

Simulations presented in \cite{2013MNRAS.430L..35G} successfully recreated observed trends in the near infrared colors of SSCs based on the evolution of the first RSG members.   RSGs contribute 90-95\% of the near-IR flux emitted by young SSCs older than $\sim$7 Myr.  Now we expand on the photometric results of \cite{2013MNRAS.430L..35G} by simulating the full spectra of SSCs as a function of age.  Here we use the same methodology as that paper to derive theoretical stellar populations of a 10$^5$ \Msun\ SSC from 1-15 Myr.  Notably, we assume a Salpeter initial mass function with mass boundaries of 0.8-100 \Msun and evolve the theoretical cluster using the Geneva stellar evolution tracks which include effects of rotation \citep{2000AA...361..101M}.  

At each time step we construct a theoretical spectral energy distribution (SED) at a resolution of R=10,000 using theoretical SEDs from the Pollux database\footnotemark[1] \citep{2010AA...516A..13P}.  This database draws from three sets of 1D LTE synthetic spectra, including {\sc cmfgen} \citep{1998ApJ...496..407H}, {\sc atlas12} using the Kurucz stellar atmospheres \citep{2005MSAIS...8...14K}, and {\sc turbospectrum} calculations using {\sc marcs} atmospheres \citep{2012ascl.soft05004P,2003ASPC..288..331G}.  For stellar parameters typical of RSGs we supplant the latter with our own \nlte\ theoretical spectra in the \jband\ \citep{2012ApJ...751..156B,2013ApJ...764..115B}.  Solar metallicity is assumed.  

\footnotetext[1]{Operated at LUPM  (UniversitŽ Montpellier II - CNRS, France) with the support of the PNPS and INSU. http://pollux.graal.univ-montp2.fr}

In Figure \ref{fig:syn1} we plot a synthetic SED from 0.3-1.5 microns at 5 Myr and 15 Myr (before and after the evolution of the first massive stars into RSGs which begins at roughly 7 Myr).  The evolution of the first RSGs have an overwhelming effect on the near IR SED, wholly dominating the flux of the cluster.  We plot two panels showing just the \jband\ in Figure \ref{fig:syn2}.  

\subsection{Analysis Tests}
 
\begin{figure}
\begin{centering}
\includegraphics[width=8.8cm]{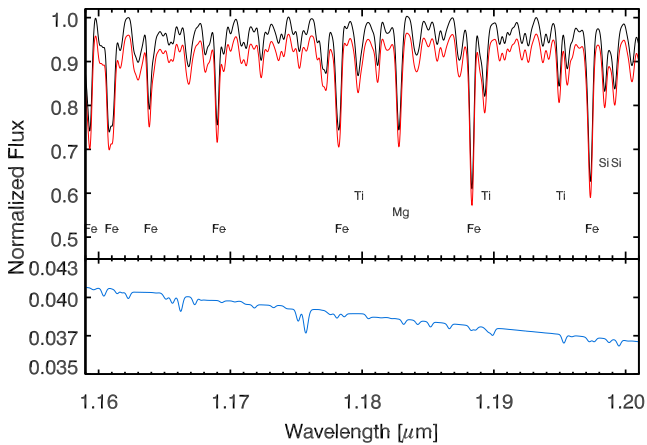}
\caption{Theoretical \jband\ spectrum of a 15 Myr old, 10$^5$ M$_\odot$ super star cluster.  Black spectrum plots the SSC spectrum normalized to unity in the \jband, red shows the contribution of the red supergiants and the lower panel provides a zoomed view of the 5-6\% level of the blue spectrum which represents the main sequence and blue supergiant stars.} 
\label{fig:syn2}
\end{centering}
\end{figure}

\begin{figure}
\begin{centering}
\includegraphics[width=8.8cm]{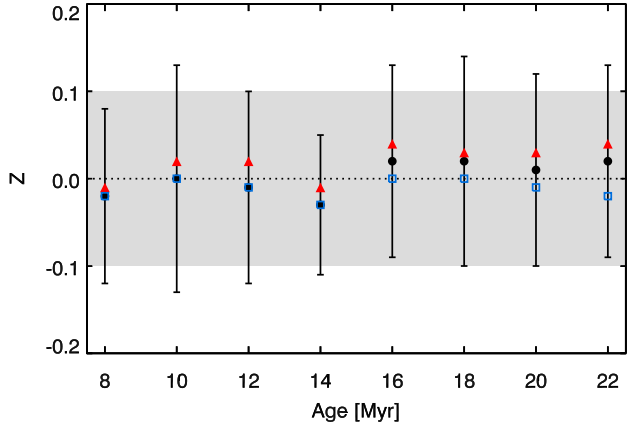}
\caption{Metallicities extracted from synthetic SSC spectra as a function of age.  Black circles show values extracted from the spectra when a 5\% flat contaminative flux is removed from each spectra , red triangles are a result of the RSG population alone, and blue squares show metallicities extracted from the full spectra.  Error bars on the black circles are consistent for each of the three types of measurement.  The gray zone is the expected region of uncertainty for a single solar metallicity red supergiant analyzed with the method used in this letter and initially presented by \cite{2010MNRAS.407.1203D} and \cite{gazak2014}.} 
\label{fig:fitt}
\end{centering}
\end{figure}

We test the hypothesis that a \jband\ spectra could yield the \met\ abundance of that cluster by applying our analysis method (presented below, \S\ref{sec:analysis}, see also \citealt{gazak2014}) to the synthetic spectra of Figure \ref{fig:syn2} between ages of 8 and 22 Myr.  For the model SSC spectra at R=3500 we recover \met\ consistent with solar metallicity and with measurement errors of 0.10 to 0.14 (See Figure \ref{fig:fitt}).  When we subtract a flat spectrum that is 5\% of the total flux (to simulate the removal of the "main sequence" contribution), the measured metallicities increase, remaining consistent with solar. 



\section{Analysis}
\label{sec:analysis}

We model the observed SSC spectra of M83-1f-117 and NGC6946-1447 (see the description of our observations and data reduction in Sec. \ref{sec:obs}) using single-star templates for synthetic spectra.  The spectra were computed as follows.
The model atmospheres are 1D LTE {\sc marcs} \citep{2008AA...486..951G}.  The coverage of the {\sc marcs} grid can be found in Table 1. Synthetic model spectra are calculated in \nlte\ for iron, titanium and silicon using the atomic models and codes described by \cite{2012ApJ...751..156B,2013ApJ...764..115B}. These atoms provide the strong lines crucial for the analysis in the \jband. The contributions by all other atoms and molecules are included in \lte.  We assume solar values for the ratios of alpha elements to iron ($\alpha$/Fe).

We begin by iteratively fitting the spectral resolution of our data by finding the best model and resolution pair by minimizing the $\chi^2$ fit statistic.  Spectral broadening due to cluster dynamics is inseparable from resolution effects. We find a resolution of R$_{\rm{eff}}$=3500 for the M83 ISAAC spectrum and R$_{\rm{eff}}$=1800 for the NGC6946 object with SpeX.  These values are consistent with the expected capability of the instruments. The measured R$_{\rm{eff}}$ is applied to the entire grid of synthetic spectra and a four-dimensional $\chi^2$ grid is calculated using the strong isolated diagnostic features of Fe\,{\sc i}, Si\,{\sc i}, and Ti\,{\sc i} across the spectral window.

Best fit parameters are extracted from the $\chi^2$ grid as follows (a detailed description of the method is given in \citealt{gazak2014}).  We construct six two-dimensional slices around the best fit model such that each $\chi^2$ slice is locked to two ``best model'' parameters and varies over the remaining.  Each slice is interpolated onto a new grid at 10$\times$ the parameter resolution.  The interpolated $\chi^2$ minimum provides a measurement of the two ``free'' parameters for each slice: over six slices we accumulate three measurements of each parameter.  The best fit values which we tabulate in Table~$\ref{tbl:par_grid}$ are the average of those three measurements.    

Standard $\chi^2$ statistics requires that the deviations between data and model in each wavelength bin be gaussian in nature.  Gaussian deviates cannot be assumed for the following reasons: the input spectrum is contaminated by other spectral types at the $\sim$5\% level, the models are likely to contain systematic errors, and residual features due to imperfect telluric corrections are not randomly normal across the spectrum.  Instead we employ a Monte Carlo test to assess the 1$\sigma$ uncertainties in our parameter extractions.  This test begins by interpolating a model to the extracted fit parameters.  We produce 1000 noise spectra as follows: generate a random gaussian deviate for each pixel such that the global standard deviation of the noise spectrum is characteristic of the signal-to-noise ratio of the observed spectrum.
We iterate over the noise spectra, adding each to the interpolated model and feeding the resulting spectrum through our fitting procedure.  Each noisy model produces a set of best fit parameters, of which the central 68\% represents a classic 1$\sigma$ region of uncertainty without assuming gaussian error processes. 

We experiment with the effect of $\sim$5-10\% contaminative flux from the remaining stellar flux of the SSC \citep{2013MNRAS.430L..35G}.  This is accomplished by assuming a flat spectral dilution and removing 5\% and 10\% of the median flux of our observed spectra.  The effect is to deepen the absorption features$-$it is in the depths of strong lines that the flat spectrum contributes the largest percent flux.  We repeat our fitting procedure after scaling out 5\% of the median flux.  These adjusted spectra yield consistent measurements of \met; measurement uncertainties dominate the shift in metallicity due to deeper absorption features.

\section{Discussion}
\label{sec:results} 

\subsection{M83}
Multiple investigations of M83's chemical enrichment using both ``direct'' and ``strong line'' \hii\ methods have produced abundance gradients across the inner and outer disk of the galaxy \citep{2002ApJ...572..838B,2005AA...441..981B,2009ApJ...695..580B}.  While plagued by the biases and uncertainties discussed in \S\ref{sec:intro}, those papers produce lower limits for the [O/H] enrichment in the inner disk of M83 of 1.6$\times$ solar and admit that the values require refinement.  In particular, \cite{2009ApJ...695..580B} find that two common calibrations of the \hii\ region method on the same dataset return identical slopes for the metallicity gradient but the measurements of the overall metallicity level vary by 0.47 dex$-$a factor of nearly three.  Furthermore, early work on \hii\ regions returned values of 2-10$\times$ solar oxygen abundance \citep{1980ApJ...236..119D}.  While current work settles around more modest values of 1.5-2$\times$, it is clear that the calibration of \hii\ region metallicities exceeding solar remains problematic.  

In Table~$\ref{tbl:par_grid2}$ we tabulate the parameters measured from our spectra corrected for a flat 5\% flux contamination.  By applying our method for extracting metallicities from the \jband\ spectra of RSGs we measure a disk metallicity of 1.9$\times$ solar (\met\ = $+$0.28 $\pm$ 0.14) for M83.  This value is consistent with \hii\ region measurements by \cite{2005AA...441..981B} who report a metallicity from \hii\  region auroral lines in the inner disk of [O/H] = 1.78$\times$ solar. 

\subsection{NGC 6946}
Measurements of the central abundance and gradient of NGC 6946 suffer from the same setbacks of the \hii\ region method.  Using two empirical calibrations, \cite{2010ApJS..190..233M} measure a central metallicity and metallicity at the isophotal radius R$_{25}$ (Z$_0$, Z$_{R_{25}}$) of 3.0$\times$ solar and 1.5$\times$ solar for the calibration of \cite{2004ApJ...617..240K} and 0.6$\times$ solar and 0.4$\times$ solar based on an alternate calibration of \cite{2005ApJ...631..231P}.  \cite{2012AA...545A..43C}, using the same two calibrations, measure Z$_0$, Z$_{R_{25}}$ of 3.4 and 1.7$\times$ solar for one and of 0.8 and 0.3$\times$ solar for the other.  In this case three of the four measured gradients are consistent but the offsets in central metallicity between calibrations are factors of four to five (0.63-0.68 dex).  We targeted the SSC NGC 6946-1447 because it has been the target of a careful, high-resolution analysis: \cite{2006MNRAS.368L..10L} use R=25,000 H and K spectra and a proprietary spectral synthesis code to measure [Fe/H] = $-$0.45 $\pm$ 0.08 (0.35$\times$ solar) and [$\alpha$/Fe] = $+$0.22 $\pm$ 0.11. 

With our method we measure a metallicity of $\sim$0.5$\times$ solar (\met\ = $-$0.32 $\pm$ 0.20).  Our measurement agrees within 1$\sigma$ to the published value in  \cite{2006MNRAS.368L..10L}, but we note that the resolution of our NGC 6946-1447 spectrum is less than ideal.  Even with this observation at modest R $\sim$1800 we can claim that the disk of this galaxy is significantly sub-solar in metallicity, something that \hii\ region methods cannot do without an arbitrary choice of calibration.  Still, the \jband\ SSC method is better suited to spectral resolutions above R=2500 (see \citealt{gazak2014}).  It is important to note that \cite{2006MNRAS.368L..10L} also find a significant enrichment in $\alpha$-elements relative to iron.  Our assumption of a solar $\alpha$/Fe will then skew our measured \met\ to higher metallicities.  Assuming the SSC does have a super-solar $\alpha$/Fe, the silicon and titanium lines in our models will be globally too shallow relative to iron.  In this case the best global fit to the spectrum using our grid will require a model with higher \met\ and may explain the difference between this work and \cite{2006MNRAS.368L..10L}.  

\subsection{Summary}

Independent techniques to measure the metallicities and gradients across the disks of star-forming galaxies are critical to our understanding of galaxy formation and evolution.  Such techniques are also poised to help disentangle the biases and poorly understood systematics inherent to ``strong line'' \hii\ methods which are routinely applied to massive datasets of galaxies.  Those techniques which have proven most successful are based on the quantitative spectroscopy of supergiant stars.  In this letter we have introduced a method capable of avoiding the extreme systematic uncertainties inherent to \hii\ region ``strong line'' methods.  We utilize the reliable quantitative spectroscopy of red supergiant stars in a new method which remains observationally efficient with existing telescopes well beyond the Local Group galaxies.  This is accomplished by targeting young super star clusters$-$coeval stellar populations dominated in the near$-$IR by red supergiant stars.   This \jband\ technique is ideally suited to multi-object R$\sim$3000 \jband\ spectrographs.  Two such instruments have recently been commissioned, KMOS on the VLT and MOSFIRE on Keck, allowing for studies of super star clusters in star-forming galaxies up to conservative distance estimates of 70 Mpc in galaxies across the northern and southern skies.  Indeed, an observation campaign is planned to push beyond this pioneering first observational step and collect spectra of super star clusters across the disk of M83 to provide an independent measurement of central metallicity and the abundance gradient of this star-forming galaxy.  

\acknowledgments
JZG and RPK acknowledge support by the National Science Foundation under grant AST-1108906 and the hospitality of the Munich University Observatory where part of this work was carried out.  BD is supported by a fellowship from the Royal Astronomical Society.  This work was partly supported by the European Union FP7 programme through ERC grant number 320360. BP is supported in part by the CNRS Programme National de Physique Stellaire. For this work the authors made use of the IRTF telescope atop Mauna Kea. The authors wish to extend special thanks to those of Hawaiian ancestry on whose sacred mountain they are privileged to be guests. Without their generous hospitality, some of the observations presented herein would not have been possible.

\end{document}